\PassOptionsToPackage{svgnames}{xcolor}
\documentclass[conference]{IEEEtran}
\usepackage[left=1.62cm,right=1.62cm,top=1.9cm]{geometry}

\usepackage[english]{babel}
\usepackage{amsmath,amssymb}
\usepackage{times}
\usepackage{relsize}
\usepackage{graphicx}
\usepackage{url}
\usepackage{booktabs}
\usepackage{multirow}
\usepackage{mparhack}
\usepackage{subfigure}
\usepackage{authblk}
\usepackage{amsthm}
\usepackage{mathrsfs}
\usepackage{pgfplots}
\pgfplotsset{width=7cm,compat=1.8}

\usepackage[noend]{algorithmic}
\usepackage[linesnumbered,ruled,vlined]{algorithm2e}
\usepackage{setspace}
\usepackage{cite}

\usepackage{footmisc}
\usepackage{array}
\usepackage{eqparbox}
\usepackage{tikz}

\usepackage{xcolor}
\usepackage{fontawesome}

\usepackage{mathtools}
\usepackage{bm}
\usepackage[all]{xy}
\usepackage{etoolbox}
\AtBeginEnvironment{figure}{\setcounter{subfigure}{0}}

\usepackage[acronym]{glossaries}
\usepgfplotslibrary{fillbetween}

\definecolor{darkgreen}{RGB}{152.9,255.0,77.9}

\SetCommentSty{mycommfont}

\usepackage{accents}

\def\I{\mathcal{I}}

\def\U{\mathcal{U}}

\def\K{\mathcal{K}}

\def\KK{\mathcal{K}^{+}}
\def\KU{\mathcal{K}^{-}}

\newacronym{csi}{CSI}{channel state information}
\newacronym{wmmse}{WMMSE}{weighted minimum mean square error}
\newacronym{sc}{SC}{synthetic control}
\newacronym{ru}{RU}{resource unit}
\newacronym{wsrm}{WSRM}{weighted sum-rate maximization}
\newacronym{ioe}{IoE}{internet of everything}
\newacronym{gnn}{GNN}{graph neural networks}
\newacronym{siso}{SISO}{single-input-single-output}
\newacronym{idi}{IDI}{interference detection and identification}
\newacronym{cca}{CCA}{clear channel assessment}
\newacronym{scm}{SCM}{structural causal model}

\renewcommand{\vec}[1]{\mathbf{#1}}

\title{Weighted Sum-Rate Maximization With Causal Inference for Latent Interference Estimation}

\author[ ]{Lei You}
\affil[ ]{\small Department of Engineering Technology, Technical University of Denmark, Denmark}
\affil[ ]{\text{leiyo@dtu.dk}}

\begin{document}

\maketitle
\thispagestyle{plain}
\pagestyle{plain}

\begin{abstract}
The paper investigates the \gls{wsrm} problem with latent interfering sources outside the known network, whose power allocation policy is hidden from and uncontrollable to optimization. The paper extends the famous alternate optimization algorithm \gls{wmmse} \cite{christensen2008weighted} under a causal inference framework to tackle with \gls{wsrm}. Specifically, 
with the possibility of power policy shifting in the hidden network, 
computing an iterating direction based only on the observed interference inherently implies that counterfactual is ignored in decision making. A method called \gls{sc} is used to estimate the counterfactual. For any link in the known network, \gls{sc} constructs a convex combination of the interference on other links and uses it as an estimate for the counterfactual. Power iteration in the proposed SC-WMMSE is performed taking into account both the observed interference and its counterfactual. SC-WMMSE requires no more information than the original WMMSE in the optimization stage. To our best knowledge, this is the first paper explores the potential of \gls{sc} in assisting mathematical optimization in addressing classic wireless optimization problems. Numerical results suggest the superiority of the SC-WMMSE over the original in both convergence and objective.
\end{abstract}
\glsresetall

\section{Introduction}
\label{sec:intro}

The wireless network evolution has been driven by a need for consistently higher rates over the past decades, along with the emergence of the \gls{ioe}, which aims to serve as a platform for connecting processes, people and data~\cite{saad2019vision}. The requirement for high spectral efficiency in \gls{ioe} scenarios calls for a need to revisit classic wireless optimization problems under highly dynamic environments. The \gls{wsrm} problem \cite{baligh2014cross} is one of these many, solving which plays a key role in determining the effective capacity of a wireless channel. 
Finding the global maximum of \gls{wsrm} is generally considered to be an $\mathcal{NP}$-hard problem~\cite{baligh2014cross}.

As a result, significant research efforts have been devoted to developing high-quality sub-optimal solutions for the problem. The \gls{wmmse} algorithm, first proposed in~\cite{christensen2008weighted}, is one such solution which has been shown to be an efficient algorithmic framework for many cross-layer transmission tasks \cite{baligh2014cross}. Therefore, this algorithmic framework has been significantly extended by many other literature and the research exploring remains still active \cite{schmidt2009minimum, shi2011iteratively, shin2012weighted, cirik2015weighted, aquilina2017weighted, guo2020weighted}. Additionally, the \gls{wsrm} problem has also been addressed by data-driven methodologies--more specifically--\gls{gnn} based methods \cite{eisen2020optimal,naderializadeh2020wireless,chowdhury2021unfolding, nikoloska2021fast}. The basic idea is to use a \gls{gnn} to encode the network topology information into a \gls{gnn} that maps \gls{csi} to power control policy. Unsupervised model training is also possible by in-cooperating learning in the processing of a primal-dual algorithm \cite{eisen2020optimal}. 

The \gls{siso} case of~\cite{schmidt2009minimum, shi2011iteratively, shin2012weighted, cirik2015weighted, aquilina2017weighted, guo2020weighted,eisen2020optimal,naderializadeh2020wireless,chowdhury2021unfolding, nikoloska2021fast} and many other research summarized in \cite{weeraddana2012weighted} falls into a special case of the problem investigated in this paper, i.e. when there are no latent interfering links whose power allocation is unknown and uncontrollable. It's important to note that \textit{the latent interfering links cannot be treated simply as noise}, because the power allocation of the latent links may be dependent on those of the known links, or, they may evolve over time. Treating it as noise omits these facts. 

The paper examines the \gls{wmmse} optimization mechanism from a causal inference's perspective. \gls{wmmse} algorithm and other similar methods can suffer performance loss in this scenario because of the influence of these latent factors on the iterating directions towards the ground-truth optimality. To overcome this challenge, the paper proposes the use of the \gls{sc} method \cite{abadie2003economic}, which uses counterfactual reasoning to infer the causal relationship between optimization variables and the objective. The resulting algorithm requires no additional information beyond that used in the traditional \gls{wmmse} algorithm, yet it demonstrates improved convergence and objective value, as well as resilience to emerging and disappearing latent interference sources. The trick in this paper hence applies to most of its previous extensions \cite{baligh2014cross,schmidt2009minimum, shi2011iteratively, shin2012weighted, cirik2015weighted, aquilina2017weighted, guo2020weighted}. Numerically, the proposed SC-WMMSE shows advantage over its origin on both convergence and objective value as well as demonstrates resilience to emerging and disappearing of latent interference sources.

\begin{figure*}[!t]
\begin{equation}
R_k = \log\bigg(1+\frac{|h_{k,k}|^2 p_k}{\sum_{j\in\KK}|h_{j,k}|^2p_j +\sum_{i\in\KU}|h_{i,k}|^2q_i+ \sigma^2_k }\bigg)\quad k\in\KK
\label{eq:KK_rate}
\end{equation}
\end{figure*}

\section{Model and Problem}

Consider a wireless network consisting of multiple communications links. The set of all links is denoted by $\K$, and for any link $k$ ($k\in\K$), the set of links that interfere with $k$ is denoted by $\I_k$, where $\I_k\subset \K$. The channel gain between the transmitter and receiver of link $k$ is denoted by $h_{k,k}$, and the channel gain from the transmitter of link $k$ to the receiver of link $j$ is denoted by $h_{k,j}$ ($k\neq j$). It is assumed that $|h_{k,j}|=0$ for any $j\notin \I_k$ ($j\neq k$), and the channel matrix is denoted by $\vec{H}$. The power of the transmitter of link $k$ is denoted by $p_k$, and the noise power is denoted by $\sigma_k^2$.



The paper considers a scenario where there are latent interference sources whose power distribution and allocation is not known a priori and is neither observable nor controllable. The known network is denoted by $\KK$, and the unknown network is denoted by $\KU$, with $\KK\cup\KU = \K$ and $\KK\cap\KU=\phi$. Let $K = |\KK|$. The two networks mutually affect each other through interference. No assumptions are made about the power policies of $\KU$. The notation $\vec{q}$ is used to represent the power allocation in $\KU$, which can proactively or reactively change over time and may be dependent on $\vec{p}$. The problem is defined in \eqref{eq:opt}, with $R_k$ defined in equation~\eqref{eq:KK_rate}. The goal is to find a power allocation $p_1,p_2,\ldots p_K$ for all the corresponding transmitters of the $K$ links in the known network such that the weighted sum rate $\sum_{k}\alpha_k R_k$ of the known network is maximized.

\begin{subequations}
\begin{align}
                 & \max_{\vec{p}} \sum_{k\in\KK}\alpha_k\mathbb{E}_{\vec{H}}\bigg[\mathbb{E}_{\vec{q}|\vec{H},\vec{p}}\big[ R_k\big] \bigg]\\
\text{s.t.}\quad & 0\leq p_{k}\leq p_k^{\max},~k\in\KK
\end{align}
\label{eq:opt}
\end{subequations}



\section{Sum Rate Maximization with Causal Inference}
\label{sec:sum-rate}


\subsection{How Latent Interference Affects \gls{wmmse}}
\label{subsec:wmmse}

It is shown that~\eqref{eq:opt} submits to a reformulation as a weighted sum-mean-square-error minimization problem \cite{schmidt2009minimum,shi2011iteratively} when  the power allocation vector of the unknown network, $\vec{q}$, is fixed as constants. When $\vec{q}$ is fixed, the mathematical expectation on $\vec{q}$ is removed from the problem.

To make this reformulation, the noise plus interference term from the unknown network $\KU$ is replaced by a new variable $\eta_k$. 
Denote 
$\eta_{k}=\sum_{i\in\KU}|h_{i,k}|^2q_i + \sigma^2_k$.
This replacement can be made in the denominator of the rate expression in \eqref{eq:KK_rate}, and the problem in \eqref{eq:opt} can be rewritten as \eqref{eq:opt_naive_reform} without loss of optimality \cite{schmidt2009minimum,shi2011iteratively}.

\begin{subequations}
\begin{align}
                 & \min_{\vec{w}, \vec{u}, \vec{v}} \mathbb{E}_{\vec{H}}\bigg[ \sum_{k\in\KK} \alpha_k(w_{k}e_k - \log w_k)\bigg] \\
\text{s.t.}  \quad& \lvert v_{k}\rvert^2 \leq p^{\max}_k,~k\in\KK
\end{align}
\label{eq:opt_naive_reform}
\end{subequations} 

The reformulation of the problem in~\eqref{eq:opt} to a weighted sum-mean-square-error minimization problem utilizes two additional variables: $w_k$ and $e_k$, where $w_k$ is a positive weight variable, and $e_k$ is the mean-square error, defined as:
\begin{equation}
e_k = |u_k h_{k,k}, v_k - 1|^2 + \sum_{j\neq k}|u_j h_{j,k}v_k|^2 +\eta_{k}|u_k|^2
\end{equation}
The variable $p_k$ in the original formulation is replaced by $v_k$, with $p_k=|v_k|^2$ ($k\in\KK$).

The \gls{wmmse} algorithm is designed to solve the reformulation in~\eqref{eq:opt_naive_reform} using the theory of alternate optimization. The algorithm is illustrated in Algorithm~\ref{alg:wmmse}. It should be noted that there also exists a stochastic version of the \gls{wmmse} algorithm that can be used when the channel matrix, $\vec{H}$, is a random variable. The proposed methodology in this paper can also be applied to the stochastic version of the \gls{wmmse} algorithm. In practice, the variable $\eta_k$ can be approximated using techniques such as \gls{cca} \cite{yang2011wireless}.


\begin{algorithm}
\setstretch{1.25}
\caption{The \gls{wmmse} algorithm for solving the problem in~\eqref{eq:opt_naive_reform}.}
\label{alg:wmmse}
\begin{algorithmic}[1]
\STATE Initialize $\vec{v}, \vec{u}, \vec{w}$ randomly
\REPEAT 
    \STATE Observe $\eta_1, \eta_2\ldots \eta_{K}$ \label{alg:1-gk}
  \FORALL{$k = 1, 2\ldots K$}
    \STATE $u_k = |h_{k,k}v_k| / (\sum_{j\in\KK}|h_{j,k}|^2|v_j|^2 + \eta_k )$ \label{alg:1-u_k}
    \STATE $w_k = 1 / (1 - |u_k h_k v_k|)$
    \STATE $v_k = \alpha_k h_k u_k w_k / (\sum_{j\in\KK}\alpha_jw_j |h_{k,j}u_j|^2 + \lambda_k)$
  \ENDFOR
\UNTIL \text{Convergence}
\RETURN $\vec{v},\vec{u},\vec{w}$
\end{algorithmic}
\end{algorithm}

The term $e_k$ is a convex quadratic function in $\vec{u}$ and $\vec{v}$, which have closed-form solutions.
However, when $\vec{q}$ is a latent random variable, it can change throughout the optimization process and affect the sampling of $\eta_k$, making the optimization process challenging. This is because the collected samples of $\eta_k$ may not guarantee that $u_k$ will move towards the optimum of the quadratic function $e_k$ when $\vec{w}$ is fixed. This problem can be mitigated if $\eta_k$ is i.i.d., yet such assumption is too strong to be realistic.

\subsection{Causal Inference Estimator vs. Regression Estimator}
\label{subsec:causal_vs_reg}
\label{subsec:opt-formulation}
The key obstacle in applying the \gls{wmmse} algorithm is that the interference at any receiver $k$ ($k\in\KK$) is difficult to handle from an optimization perspective, due to its dependency on $\vec{q}$. To tackle this obstacle, instead of considering only $\eta_{k}$, the entire denominator of $u_k$, denoted by $I_{k}$, is considered. This allows for better generalization of the proposed inference methodology, as it can estimate the denominator as a whole for any link $k$ if the \gls{csi} of another link $j$ ($j\neq k$, $j\in\KK$) is unknown or outdated. In the $k_{\text{th}}$ loop of Algorithm~\ref{alg:wmmse}, the power allocation $p_k$ is the variable to be optimized and the other power $\vec{p}_{-k}$ are fixed. The mathematical expectation of $I_k$ conditional on the optimization variable $p_k$ is as follows.
\begin{equation*}
\begin{split}
   \MoveEqLeft  \mathbb{E}[I_{k} | p_k] \\ 
    \!\!={}& \mathbb{E}_{\vec{H}}\mathbb{E}_{\vec{q}|\vec{H},p_k}\bigg[\sum_{j\in\KK}\!\!|h_{j,k}|^2p_j + \sum_{i\in\KU}\!\!|h_{i,k}|^2q_i + \sigma_k^2\bigg]  \\
    \!\!={}& \mathbb{E}_{\vec{H}}\bigg[\!\sum_{j\in\KK}\!\!|h_{j,k}|^2p_j\bigg] + \underbrace{\mathbb{E}_{\vec{H}}\mathbb{E}_{\vec{q}|\vec{H},p_k}\bigg[\!\sum_{i\in\KU}\!\!\!|h_{i,k}|^2q_i+\sigma_k^2\bigg]}_{\eta_{k}}
\end{split}
\end{equation*}

Estimating the second part of $I_k$, $\eta_k$, is difficult as it depends on the power allocation of the latent interfering sources, which is unknown and may change over time. Supervised machine learning models are limited for this task as the distribution of $\vec{q}$ is not i.i.d and may change due to changes in the power allocation of the latent sources or new transmitters/receivers joining or leaving the network. A more suitable approach is to leverage causal inference methods, specifically the \gls{scm} approach, to inexplicitly estimate the counterfactual distribution of $\vec{q}$ given the observed interference $I_k$. This approach allows for estimation of the interference distribution under different power allocation policies, which can then be used to infer the causality relationship between $p_k$ and $I_k$. 

 A causal inference estimator  targets $\mathbb{E}[I_k|do(p_k)]$, which takes both the actual outcome and potential outcomes into consideration, rather than just the former. Specifically, $\mathbb{E}[I_k|do(p_k)]$ represents the expected value of the interference $I_k$ at receiver $k$ when we intervene and set the power allocation of link $k$ to $p_k$, and consider all possible power policies of the latent interfering links $\KU$. On the other hand, $\mathbb{E}[I_k|p_k]$ represents the expected value of the interference $I_k$ at receiver $k$ when the power allocation of link $k$ is $p_k$, but the power policies of the latent interfering links $\KU$ are not intervened, and are instead determined by the current distribution of $\vec{q}$ given $\vec{H}$ and $\vec{p}$. $\mathbb{E}[I_k|do(p_k)]$ better aligns with the requirements of optimization.
 
For $\mathbb{E}[I_k|do(p_k)]$ and $\mathbb{E}[I_k|p_k]$ to be equal, the power allocation $p_k$ is independent of all the other interfering links and their power allocations in the network, i.e., there are no confounding variables. In this case, conditioning on $p_k$ would not change the distribution of $I_k$ and the expectation would be the same whether we condition on $p_k$ or intervene on it. This implies that power allocation in $\KK$ is being performed randomly, rather than through a specific optimization algorithm.

\subsection{Estimating Interference with \gls{sc} Methods}
\label{subsec:sc}

\begin{figure}
\centering
\tikzset{every picture/.style={line width=0.75pt}} 

\begin{tikzpicture}[x=0.75pt,y=0.75pt,yscale=-1,xscale=1]

\draw   (366.92,228.47) -- (320.13,292.88) -- (244.42,268.28) -- (244.42,188.67) -- (320.13,164.07) -- cycle ;
\draw  [color={rgb, 255:red, 208; green, 2; blue, 27 }  ,draw opacity=1 ][fill={rgb, 255:red, 208; green, 2; blue, 27 }  ,fill opacity=1 ] (364.01,179.62) .. controls (364.01,177.61) and (365.65,175.98) .. (367.66,175.98) .. controls (369.67,175.98) and (371.3,177.61) .. (371.3,179.62) .. controls (371.3,181.63) and (369.67,183.26) .. (367.66,183.26) .. controls (365.65,183.26) and (364.01,181.63) .. (364.01,179.62) -- cycle ;
\draw [color={rgb, 255:red, 74; green, 144; blue, 226 }  ,draw opacity=1 ][line width=0.75]  [dash pattern={on 4.5pt off 4.5pt}]  (322.45,206.36) -- (366.92,228.47) ;
\draw [color={rgb, 255:red, 74; green, 144; blue, 226 }  ,draw opacity=1 ][line width=0.75]  [dash pattern={on 4.5pt off 4.5pt}]  (322.45,206.36) -- (320.13,292.88) ;
\draw [color={rgb, 255:red, 74; green, 144; blue, 226 }  ,draw opacity=1 ][line width=0.75]  [dash pattern={on 4.5pt off 4.5pt}]  (244.42,268.28) -- (322.45,206.36) ;
\draw [color={rgb, 255:red, 74; green, 144; blue, 226 }  ,draw opacity=1 ][line width=0.75]  [dash pattern={on 4.5pt off 4.5pt}]  (322.45,206.36) -- (244.42,188.67) ;
\draw [color={rgb, 255:red, 208; green, 2; blue, 27 }  ,draw opacity=1 ][line width=0.75]  [dash pattern={on 4.5pt off 4.5pt}]  (320.13,292.88) -- (367.66,179.62) ;
\draw [color={rgb, 255:red, 208; green, 2; blue, 27 }  ,draw opacity=1 ][line width=0.75]  [dash pattern={on 4.5pt off 4.5pt}]  (244.42,268.28) -- (367.66,179.62) ;
\draw [color={rgb, 255:red, 208; green, 2; blue, 27 }  ,draw opacity=1 ][line width=0.75]  [dash pattern={on 4.5pt off 4.5pt}]  (367.66,179.62) -- (366.92,228.47) ;
\draw [color={rgb, 255:red, 208; green, 2; blue, 27 }  ,draw opacity=1 ][line width=0.75]  [dash pattern={on 4.5pt off 4.5pt}]  (364.01,179.62) -- (320.13,164.07) ;
\draw  [fill={rgb, 255:red, 0; green, 0; blue, 0 }  ,fill opacity=1 ] (363.28,228.47) .. controls (363.28,226.46) and (364.91,224.83) .. (366.92,224.83) .. controls (368.93,224.83) and (370.56,226.46) .. (370.56,228.47) .. controls (370.56,230.49) and (368.93,232.12) .. (366.92,232.12) .. controls (364.91,232.12) and (363.28,230.49) .. (363.28,228.47) -- cycle ;
\draw  [fill={rgb, 255:red, 0; green, 0; blue, 0 }  ,fill opacity=1 ] (316.49,164.07) .. controls (316.49,162.06) and (318.12,160.43) .. (320.13,160.43) .. controls (322.14,160.43) and (323.77,162.06) .. (323.77,164.07) .. controls (323.77,166.09) and (322.14,167.72) .. (320.13,167.72) .. controls (318.12,167.72) and (316.49,166.09) .. (316.49,164.07) -- cycle ;
\draw  [fill={rgb, 255:red, 0; green, 0; blue, 0 }  ,fill opacity=1 ] (240.78,188.67) .. controls (240.78,186.66) and (242.41,185.03) .. (244.42,185.03) .. controls (246.43,185.03) and (248.07,186.66) .. (248.07,188.67) .. controls (248.07,190.68) and (246.43,192.32) .. (244.42,192.32) .. controls (242.41,192.32) and (240.78,190.68) .. (240.78,188.67) -- cycle ;
\draw  [fill={rgb, 255:red, 0; green, 0; blue, 0 }  ,fill opacity=1 ] (240.78,268.28) .. controls (240.78,266.26) and (242.41,264.63) .. (244.42,264.63) .. controls (246.43,264.63) and (248.07,266.26) .. (248.07,268.28) .. controls (248.07,270.29) and (246.43,271.92) .. (244.42,271.92) .. controls (242.41,271.92) and (240.78,270.29) .. (240.78,268.28) -- cycle ;
\draw  [fill={rgb, 255:red, 0; green, 0; blue, 0 }  ,fill opacity=1 ] (316.49,292.88) .. controls (316.49,290.86) and (318.12,289.23) .. (320.13,289.23) .. controls (322.14,289.23) and (323.77,290.86) .. (323.77,292.88) .. controls (323.77,294.89) and (322.14,296.52) .. (320.13,296.52) .. controls (318.12,296.52) and (316.49,294.89) .. (316.49,292.88) -- cycle ;
\draw  [color={rgb, 255:red, 74; green, 144; blue, 226 }  ,draw opacity=1 ][fill={rgb, 255:red, 74; green, 144; blue, 226 }  ,fill opacity=1 ] (318.8,206.36) .. controls (318.8,204.35) and (320.43,202.72) .. (322.45,202.72) .. controls (324.46,202.72) and (326.09,204.35) .. (326.09,206.36) .. controls (326.09,208.38) and (324.46,210.01) .. (322.45,210.01) .. controls (320.43,210.01) and (318.8,208.38) .. (318.8,206.36) -- cycle ;
\draw [line width=0.75]    (190.8,324.49) -- (434.77,324.49) ;
\draw [shift={(436.77,324.49)}, rotate = 180] [color={rgb, 255:red, 0; green, 0; blue, 0 }  ][line width=0.75]    (10.93,-3.29) .. controls (6.95,-1.4) and (3.31,-0.3) .. (0,0) .. controls (3.31,0.3) and (6.95,1.4) .. (10.93,3.29)   ;
\draw [line width=0.75]    (204.52,339.19) -- (204.52,129.52) ;
\draw [shift={(204.52,127.52)}, rotate = 90] [color={rgb, 255:red, 0; green, 0; blue, 0 }  ][line width=0.75]    (10.93,-3.29) .. controls (6.95,-1.4) and (3.31,-0.3) .. (0,0) .. controls (3.31,0.3) and (6.95,1.4) .. (10.93,3.29)   ;
\draw [color={rgb, 255:red, 208; green, 2; blue, 27 }  ,draw opacity=1 ][line width=0.75]  [dash pattern={on 4.5pt off 4.5pt}]  (364.01,179.62) -- (248.07,188.67) ;
\draw [color={rgb, 255:red, 74; green, 144; blue, 226 }  ,draw opacity=1 ][line width=0.75]  [dash pattern={on 4.5pt off 4.5pt}]  (322.45,206.36) -- (320.13,167.72) ;

\draw (323.23,149.49) node [anchor=north west][inner sep=0.75pt]  [font=\normalsize]  {$I_{1}$};
\draw (369.25,217.36) node [anchor=north west][inner sep=0.75pt]  [font=\normalsize]  {$I_{2}$};
\draw (322.86,285.22) node [anchor=north west][inner sep=0.75pt]  [font=\normalsize]  {$I_{3}$};
\draw (231.06,266.42) node [anchor=north west][inner sep=0.75pt]  [font=\normalsize]  {$I_{4}$};
\draw (225.1,177.92) node [anchor=north west][inner sep=0.75pt]  [font=\normalsize]  {$I_{5}$};
\draw (319.55,182.82) node [anchor=north west][inner sep=0.75pt]  [font=\normalsize]  {$\hat{I}_{6}$};
\draw (370.06,161.75) node [anchor=north west][inner sep=0.75pt]  [font=\normalsize]  {$\hat{I}_{6}$};
\draw (422.83,329.15) node [anchor=north west][inner sep=0.75pt]    {$p_{x}$};
\draw (176.86,121.4) node [anchor=north west][inner sep=0.75pt]    {$p_{y}$};
\draw (190.14,323.58) node [anchor=north west][inner sep=0.75pt]    {$0$};

\end{tikzpicture}
\caption{Illustration of interpolation vs. extrapolation. Consider $\KK=\{1,2,3,4,5,6\}$ and $\KU=\{x,y\}$. Note that $q_x$ and $q_y$ are latent variables to $\KK$, influencing the observed interference at each link of $\KK$. The positions of $I_1,I_2\ldots I_5$ are affected by the value of $q_x$ and $q_y$ at the moment of observation. Estimations are performed for the link $6$, denoted by $\hat{I}_6$ with blue and red dots. The blue is from intertropolation, as it falls in-between the known observations (i.e. the convex hull), whereas the red is from extrapolation. }
\label{fig: constrained_lr}
\end{figure}
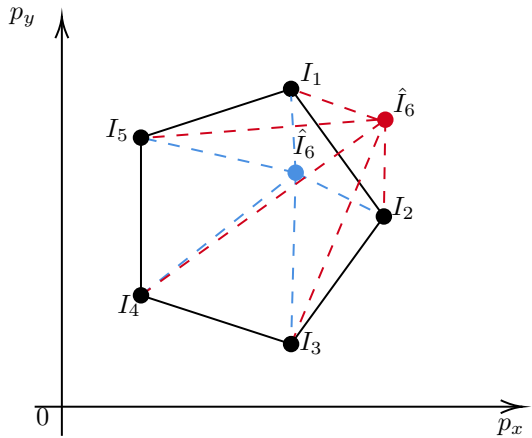

The synthetic control method (SC), first proposed in \cite{abadie2003economic}, is a powerful technique for estimating the effects of large-scale interventions \cite{shi2022assumptions}. It has been shown to be competitive with other fixed matching estimators \cite{chen2022synthetic}. SC approximates the counterfactual outcomes of one unit by constructing a weighted combination of the observed outcomes of other units. It works with panel data, which contains multiple observations for each unit over time. While SC is often used in the context of binary treatment and no interference between units, it can also be applied in more complex settings with units interfering each others, such as in \cite{agarwal2022network}, where the potential outcome for unit $k\in\KK$ is linear to latent factors, with noise in the factor model being additive, zero mean, and independent.

In the training stage, consider observations of $I_1,I_2\ldots I_k$ in format of panel data, i.e.
\begin{equation*}
\vec{X}=
\begin{bmatrix}
I_1^{(0)} & I_2^{(0)} & \cdots  & I^{(0)}_{K}  \\
I_1^{(1)} & I_2^{(1)} & \cdots  & I^{(1)}_{K}  \\
 \vdots & \vdots  & \ddots & \vdots      \\
I_1^{(L)} & I_2^{(L)} & \cdots  & I^{(L)}_{K}  
\end{bmatrix} \in \mathbb{R}^{L\times K}
\end{equation*}
where each row $\ell$ ($1\leq\ell\leq L$) is an observation across all the $K$ units. For any $k\in\KK$, denote by $\vec{x}_k$ the $k_{\text{th}}$ column of $X$. Denote by $\vec{X}_{-k}$ the matrix without column $k$. The \gls{sc} estimator is trained by solving the constrained optimization problem~\eqref{eq:sc} as follows.
\begin{subequations}
\begin{align}
                 & \bm{\nu}_k = \arg\min_{\bm{\beta}\geq\vec{0}} \| \vec{x}_{k} - \vec{X}_{-k}\bm{\beta}\| \label{eq:sc-obj} \\
\text{s.t.}  \quad& \sum_{i} \beta_i = 1   \quad i=1,2\ldots K-1 \label{eq:sc-constr}
\end{align}
\label{eq:sc}
\end{subequations}

Solving \eqref{eq:sc} yields a vector $\bm{\nu}_{k}$ ($k\in\KK$), which is essentially a group of coefficients that can be used to construct a linear combination of $I_j$ ($j\in\KK\backslash {k}$). The objective function~\eqref{eq:sc-obj} suggests that the computed coefficients lead to an as small as possible mean-squared-error over all the $L$ observations of the unit $k$ and respectively its constructed linear combinations. Note that \eqref{eq:sc-constr} imposes a hard constraint on the coefficients such that the obtained linear combination is guaranteed to be a convex combination. In other words, solving the formulation~\eqref{eq:sc} trains a (constrained) linear regression model between the observed interference of unit $k$ and those of the others. In this context, a convex combination has better interpretability. Namely, in terms of interference, it finds a function that accurately estimates the values of the counterfactual interference in the range of the observed interference, as opposed to extrapolation. See \figurename~\ref{fig: constrained_lr} for an illustration. Section~\ref{sec:simulation} demonstrates the necessity of constraint~\eqref{eq:sc-constr} by showing numerically that it helps improve optimization significantly.

In the inference stage, the synthetic control method is used to estimate the counterfactual outcome for unit $k$ under treatment $p_k$. This is done by averaging the outcomes of the synthetic control units ($\KK\backslash\{k\}$), weighted by the coefficients obtained in the training stage. The estimation is denoted as $\hat{I}_k = \bm{\nu_}k^{\intercal}\bm{\mu}_k$ for all $k\in\KK$, where $\bm{\mu}_{k}$ denotes the observations for all units other than $k$, i.e. any $I_j$ ($j\neq k$). 

\subsection{Algorithm Design}
\label{subsec:sc-wmmse}

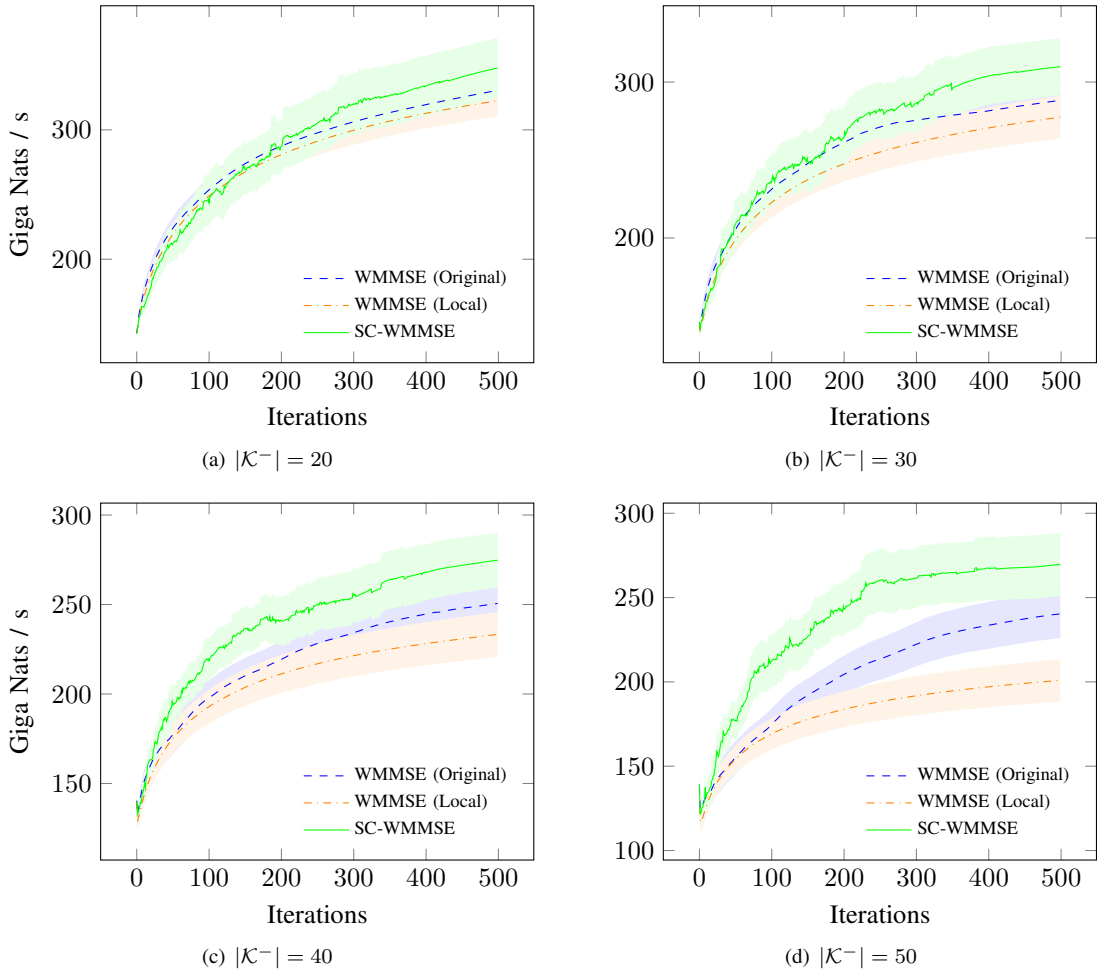
\begin{figure*}[t]
\centering

\subfigure[$|\KU|=20$\label{fig:dep-dep-k20}]{
\begin{tikzpicture}
\begin{axis}[
    ymin=120, 
    xlabel={Iterations},
    ylabel={Giga Nats / s},
    legend style={font=\scriptsize,draw=none, fill=none},
    legend pos=south east,
    legend cell align={left},
    width = 0.4\textwidth,
]
\addplot [blue, dashed] table [x=iteration, y=wmmse]{fig1-1.txt};
\addplot [orange, dash dot] table [x=iteration, y=wmmse_local]{fig1-1.txt};
\addplot [green] table [x=iteration, y=wmmse_sc]{fig1-1.txt};
\legend{WMMSE (Original), WMMSE (Local), SC-WMMSE}

\addplot [name path=lower, fill=none, draw=none] table [
    x=iteration, y expr=\thisrow{wmmse} - \thisrow{wmmse_ci}]{fig1-1.txt};
\addplot [name path=upper, fill=none, draw=none] table [
    x=iteration, y expr=\thisrow{wmmse} + \thisrow{wmmse_ci}]{fig1-1.txt};
\addplot[blue!10] fill between[of=lower and upper];

\addplot [name path=lower, fill=none, draw=none] table [
    x=iteration, y expr=\thisrow{wmmse_local} - \thisrow{wmmse_local_ci}]{fig1-1.txt};
\addplot [name path=upper, fill=none, draw=none] table [
    x=iteration, y expr=\thisrow{wmmse_local} + \thisrow{wmmse_local_ci}]{fig1-1.txt};
\addplot[orange!10] fill between[of=lower and upper];

\addplot [name path=lower, fill=none, draw=none] table [
    x=iteration, y expr=\thisrow{wmmse_sc} - \thisrow{wmmse_sc_ci}]{fig1-1.txt};
\addplot [name path=upper, fill=none, draw=none] table [
    x=iteration, y expr=\thisrow{wmmse_sc} + \thisrow{wmmse_sc_ci}]{fig1-1.txt};
\addplot[green!10] fill between[of=lower and upper];

\end{axis}
\end{tikzpicture}
}\quad\quad
\subfigure[$|\KU|=30$\label{fig:dep-dep-k30}]{
\begin{tikzpicture}
\begin{axis}[
    ymin=120, 
    xlabel={Iterations},
    legend style={font=\scriptsize,draw=none, fill=none},
    legend pos=south east,
    legend cell align={left},
    width = 0.4\textwidth,
]
\addplot [blue, dashed] table [x=iteration, y=wmmse]{fig1-2.txt};
\addplot [orange, dash dot] table [x=iteration, y=wmmse_local]{fig1-2.txt};
\addplot [green] table [x=iteration, y=wmmse_sc]{fig1-2.txt};
\legend{WMMSE (Original), WMMSE (Local), SC-WMMSE}

\addplot [name path=lower, fill=none, draw=none] table [
    x=iteration, y expr=\thisrow{wmmse} - \thisrow{wmmse_ci}]{fig1-2.txt};
\addplot [name path=upper, fill=none, draw=none] table [
    x=iteration, y expr=\thisrow{wmmse} + \thisrow{wmmse_ci}]{fig1-2.txt};
\addplot[blue!10] fill between[of=lower and upper];

\addplot [name path=lower, fill=none, draw=none] table [
    x=iteration, y expr=\thisrow{wmmse_local} - \thisrow{wmmse_local_ci}]{fig1-2.txt};
\addplot [name path=upper, fill=none, draw=none] table [
    x=iteration, y expr=\thisrow{wmmse_local} + \thisrow{wmmse_local_ci}]{fig1-2.txt};
\addplot[orange!10] fill between[of=lower and upper];

\addplot [name path=lower, fill=none, draw=none] table [
    x=iteration, y expr=\thisrow{wmmse_sc} - \thisrow{wmmse_sc_ci}]{fig1-2.txt};
\addplot [name path=upper, fill=none, draw=none] table [
    x=iteration, y expr=\thisrow{wmmse_sc} + \thisrow{wmmse_sc_ci}]{fig1-2.txt};
\addplot[green!10] fill between[of=lower and upper];

\end{axis}
\end{tikzpicture}
}

\subfigure[$|\KU|=40$\label{fig:dep-dep-k40}]{
\begin{tikzpicture}
\begin{axis}[
    xlabel={Iterations},
    ylabel={Giga Nats / s},
    legend style={font=\scriptsize,draw=none, fill=none},
    legend pos=south east,
    legend cell align={left},
    width = 0.4\textwidth,
]
\addplot [blue, dashed] table [x=iteration, y=wmmse]{fig1-3.txt};
\addplot [orange, dash dot] table [x=iteration, y=wmmse_local]{fig1-3.txt};
\addplot [green] table [x=iteration, y=wmmse_sc]{fig1-3.txt};
\legend{WMMSE (Original), WMMSE (Local), SC-WMMSE}

\addplot [name path=lower, fill=none, draw=none] table [
    x=iteration, y expr=\thisrow{wmmse} - \thisrow{wmmse_ci}]{fig1-3.txt};
\addplot [name path=upper, fill=none, draw=none] table [
    x=iteration, y expr=\thisrow{wmmse} + \thisrow{wmmse_ci}]{fig1-3.txt};
\addplot[blue!10] fill between[of=lower and upper];

\addplot [name path=lower, fill=none, draw=none] table [
    x=iteration, y expr=\thisrow{wmmse_local} - \thisrow{wmmse_local_ci}]{fig1-3.txt};
\addplot [name path=upper, fill=none, draw=none] table [
    x=iteration, y expr=\thisrow{wmmse_local} + \thisrow{wmmse_local_ci}]{fig1-3.txt};
\addplot[orange!10] fill between[of=lower and upper];

\addplot [name path=lower, fill=none, draw=none] table [
    x=iteration, y expr=\thisrow{wmmse_sc} - \thisrow{wmmse_sc_ci}]{fig1-3.txt};
\addplot [name path=upper, fill=none, draw=none] table [
    x=iteration, y expr=\thisrow{wmmse_sc} + \thisrow{wmmse_sc_ci}]{fig1-3.txt};
\addplot[green!10] fill between[of=lower and upper];

\end{axis}
\end{tikzpicture}
}\quad\quad
\subfigure[$|\KU|=50$\label{fig:dep-dep-k50}]{
\begin{tikzpicture}
\begin{axis}[
    xlabel={Iterations},
    legend style={font=\scriptsize,draw=none, fill=none},
    legend pos=south east,
    legend cell align={left},
    width = 0.4\textwidth,
]
\addplot [blue, dashed] table [x=iteration, y=wmmse]{fig1-4.txt};
\addplot [orange, dash dot] table [x=iteration, y=wmmse_local]{fig1-4.txt};
\addplot [green] table [x=iteration, y=wmmse_sc]{fig1-4.txt};
\legend{WMMSE (Original), WMMSE (Local), SC-WMMSE}

\addplot [name path=lower, fill=none, draw=none] table [
    x=iteration, y expr=\thisrow{wmmse} - \thisrow{wmmse_ci}]{fig1-4.txt};
\addplot [name path=upper, fill=none, draw=none] table [
    x=iteration, y expr=\thisrow{wmmse} + \thisrow{wmmse_ci}]{fig1-4.txt};
\addplot[blue!10] fill between[of=lower and upper];

\addplot [name path=lower, fill=none, draw=none] table [
    x=iteration, y expr=\thisrow{wmmse_local} - \thisrow{wmmse_local_ci}]{fig1-4.txt};
\addplot [name path=upper, fill=none, draw=none] table [
    x=iteration, y expr=\thisrow{wmmse_local} + \thisrow{wmmse_local_ci}]{fig1-4.txt};
\addplot[orange!10] fill between[of=lower and upper];

\addplot [name path=lower, fill=none, draw=none] table [
    x=iteration, y expr=\thisrow{wmmse_sc} - \thisrow{wmmse_sc_ci}]{fig1-4.txt};
\addplot [name path=upper, fill=none, draw=none] table [
    x=iteration, y expr=\thisrow{wmmse_sc} + \thisrow{wmmse_sc_ci}]{fig1-4.txt};
\addplot[green!10] fill between[of=lower and upper];

\end{axis}
\end{tikzpicture}
}
\caption{The performance in objective maximization and convergence is evaluated over $50$ independent simulations. The 90\% confidence interval is plotted for each curve. The setup $|\KK|=50$ is used throughout \ref{fig:dep-dep-k20}--\ref{fig:dep-dep-k50}. The y-axis is the sum-rate over all links in $\KK$.}
\label{fig:convergence}
\end{figure*}

The algorithm SC-WMMSE is designed straightforwardly by incorporating the ideas discussed in Sections~\ref{subsec:wmmse}, \ref{subsec:causal_vs_reg}, and \ref{subsec:sc}. It is presented in Algorithm~\ref{alg:sc-wmmse}.

\begin{algorithm}
\setstretch{1.25}
\caption{The SC-WMMSE algorithm for solving the problem in~\eqref{eq:opt}}
\label{alg:sc-wmmse}
\begin{algorithmic}[1]
\STATE Initialize $\vec{v}, \vec{u}, \vec{w}$ randomly
\STATE Train estimators $\bm{\nu}_1, \bm{\nu}_2\ldots \bm{\nu}_{K}$ offline by~\eqref{eq:sc} \label{alg:sc-wmmse-train}
\REPEAT 
    \STATE Observe $I_1, I_2, \ldots I_{K}$ under $\vec{v}$
  \FORALL{$k = 1, 2\ldots K$}
    \STATE  $\bm{\mu}_{k}=[I_1,\ldots I_{k-1}, I_{k+1},\ldots I_{K}]$
    \IF{\textit{Rand()} $<\varepsilon$}\label{alg:sc-wmmse-if}
        \STATE $u_k = |h_{k,k}v_k|/\bm{\nu}_k^{\top}\bm{\mu}_k$ \tcp{Counterfactual update} \label{alg:sc-wmmse-estimate}
    \ELSE 
         \STATE $u_k = |h_{k,k}v_k|/I_k$ \tcp{Factual update} \label{alg:sc-wmmse-observed}
    \ENDIF
    \STATE $w_k = 1 / (1 - |u_k h_k v_k|)$
    \STATE $v_k = \alpha_k h_k u_k w_k / (\sum_{j\in\KK}\alpha_jw_j |h_{k,j}u_j|^2 + \lambda_k)$
  \ENDFOR
\UNTIL \text{Convergence}
\RETURN $\vec{p} = \big[v^2_1, v^2_2,\ldots v^2_{K}\big]$
\end{algorithmic}
\end{algorithm}

The algorithm SC-WMMSE combines the \gls{wmmse} algorithm with \gls{sc} method, which is designed to estimate the counterfactual outcome of a unit under a certain treatment. The algorithm is trained offline using past observations of $I_1,I_2,\ldots I_K$ and $K$ \gls{sc} estimators are trained. During optimization, $I_1,I_2,\ldots I_K$ are observed in every iteration and the \gls{sc} estimators are used to update $u_k$ with a probability $\varepsilon$, otherwise, the update follows the same rule as WMMSE. The parameter $\varepsilon$ is set to decay based on the number of iterations for the convergence of the algorithm\footnote{In the implementation of this paper, the formula $\varepsilon(t)=[a(1-t/t^{\max})]^{b}$ is used, where $t$ is the iteration index and $a$, $b$ are hyper-parameters. As for this paper, the setting $a=0.2$ and $b=2$ stays unchanged throughout all simulations in Section~\ref{sec:simulation}.}. The effectiveness of the algorithm is demonstrated in the simulation section of the paper. 

\subsection{Intuition Behind}
From a causal inference perspective, the optimization process of \gls{wmmse} makes treatment decision $p_k$ by its knowledge of how such an intervene would affect the outcome $I_k$ at each loop $k$. The covariates are $\vec{q}$, $\vec{H}$ (which are related to network topology and density). The \gls{sc} method is used to estimate the impact of this treatment by comparing the treated group ($\{k\}$) to a \gls{sc} group, which is created from a weighted combination of untreated groups ($\KK\backslash\{k\}$).

Ideally, the \gls{sc} group ought to be selected carefully to closely mimic the characteristics of the treated group before the treatment is applied. Optimizing the selection of the \gls{sc} group is out of the scope of this paper. The \gls{sc} method allows us to estimate what would have happened to the link $k$ if $p_k$ had not been applied, making it a counterfactual analysis. In Algorithm~\ref{alg:sc-wmmse}, line~\ref{alg:sc-wmmse-estimate} can be viewed as a "counterfactual update" and line~\ref{alg:sc-wmmse-observed} as a regular update. The optimization variable $u_k$ is therefore updated using both the observation and the counterfactual, allowing us to address cases where some observed $I_k$ are not statistically caused by the treatment $p_k$, but rather by other pre-treatment characteristics.

It is worth noting that the power allocation serves both as the intervention of causal inference and the optimization for sum-rate maximization. How to best utilize this interplay remains an open area of study.

\begin{figure}
\centering
\subfigure[$|\KU|=0$ at both the training and the inference stages\label{fig:zero-zero}]{
\begin{tikzpicture}
\begin{axis}[
    xlabel={Iterations},
    ylabel={Giga Nats / s},
    legend style={font=\scriptsize,draw=none, fill=none},
    legend pos=south east,
    legend cell align={left},
    width = 0.4\textwidth,
]
\addplot [blue, dashed] table [x=iteration, y=wmmse]{fig2-1.txt};
\addplot [orange, dash dot] table [x=iteration, y=wmmse_local]{fig2-1.txt};
\addplot [green] table [x=iteration, y=wmmse_sc]{fig2-1.txt};
\legend{WMMSE (Original), WMMSE (Local), SC-WMMSE}

\addplot [name path=lower, fill=none, draw=none] table [
    x=iteration, y expr=\thisrow{wmmse} - \thisrow{wmmse_ci}]{fig2-1.txt};
\addplot [name path=upper, fill=none, draw=none] table [
    x=iteration, y expr=\thisrow{wmmse} + \thisrow{wmmse_ci}]{fig2-1.txt};
\addplot[blue!10] fill between[of=lower and upper];

\addplot [name path=lower, fill=none, draw=none] table [
    x=iteration, y expr=\thisrow{wmmse_local} - \thisrow{wmmse_local_ci}]{fig2-1.txt};
\addplot [name path=upper, fill=none, draw=none] table [
    x=iteration, y expr=\thisrow{wmmse_local} + \thisrow{wmmse_local_ci}]{fig2-1.txt};
\addplot[orange!10] fill between[of=lower and upper];

\addplot [name path=lower, fill=none, draw=none] table [
    x=iteration, y expr=\thisrow{wmmse_sc} - \thisrow{wmmse_sc_ci}]{fig2-1.txt};
\addplot [name path=upper, fill=none, draw=none] table [
    x=iteration, y expr=\thisrow{wmmse_sc} + \thisrow{wmmse_sc_ci}]{fig2-1.txt};
\addplot[green!10] fill between[of=lower and upper];

\end{axis}
\end{tikzpicture}
}

\subfigure[$|\KU|=50$ at training. $|\KU|=0$ at inference.\label{fig:dependent-zero}]{
\begin{tikzpicture}
\begin{axis}[
    xlabel={Iterations},
    ylabel={Giga Nats / s},
    legend style={font=\scriptsize,draw=none, fill=none},
    legend pos=south east,
    legend cell align={left},
    width = 0.4\textwidth,
]
\addplot [blue, dashed] table [x=iteration, y=wmmse]{fig2-2.txt};
\addplot [orange, dash dot] table [x=iteration, y=wmmse_local]{fig2-2.txt};
\addplot [green] table [x=iteration, y=wmmse_sc]{fig2-2.txt};
\legend{WMMSE (Original), WMMSE (Local), SC-WMMSE}

\addplot [name path=lower, fill=none, draw=none] table [
    x=iteration, y expr=\thisrow{wmmse} - \thisrow{wmmse_ci}]{fig2-2.txt};
\addplot [name path=upper, fill=none, draw=none] table [
    x=iteration, y expr=\thisrow{wmmse} + \thisrow{wmmse_ci}]{fig2-2.txt};
\addplot[blue!10] fill between[of=lower and upper];

\addplot [name path=lower, fill=none, draw=none] table [
    x=iteration, y expr=\thisrow{wmmse_local} - \thisrow{wmmse_local_ci}]{fig2-2.txt};
\addplot [name path=upper, fill=none, draw=none] table [
    x=iteration, y expr=\thisrow{wmmse_local} + \thisrow{wmmse_local_ci}]{fig2-2.txt};
\addplot[orange!10] fill between[of=lower and upper];

\addplot [name path=lower, fill=none, draw=none] table [
    x=iteration, y expr=\thisrow{wmmse_sc} - \thisrow{wmmse_sc_ci}]{fig2-2.txt};
\addplot [name path=upper, fill=none, draw=none] table [
    x=iteration, y expr=\thisrow{wmmse_sc} + \thisrow{wmmse_sc_ci}]{fig2-2.txt};
\addplot[green!10] fill between[of=lower and upper];

\end{axis}
\end{tikzpicture}
}

\subfigure[$|\KU|=0$ at training. $|\KU|=50$ at inference. \label{fig:zero-dependent}]{
\begin{tikzpicture}
\begin{axis}[
    xlabel={Iterations},
    ylabel={Giga Nats / s},
    legend style={font=\scriptsize,draw=none, fill=none},
    legend pos=south east,
    legend cell align={left},
    width = 0.4\textwidth,
]
\addplot [blue, dashed] table [x=iteration, y=wmmse]{fig2-3.txt};
\addplot [orange, dash dot] table [x=iteration, y=wmmse_local]{fig2-3.txt};
\addplot [green] table [x=iteration, y=wmmse_sc]{fig2-3.txt};
\legend{WMMSE (Original), WMMSE (Local), SC-WMMSE}

\addplot [name path=lower, fill=none, draw=none] table [
    x=iteration, y expr=\thisrow{wmmse} - \thisrow{wmmse_ci}]{fig2-3.txt};
\addplot [name path=upper, fill=none, draw=none] table [
    x=iteration, y expr=\thisrow{wmmse} + \thisrow{wmmse_ci}]{fig2-3.txt};
\addplot[blue!10] fill between[of=lower and upper];

\addplot [name path=lower, fill=none, draw=none] table [
    x=iteration, y expr=\thisrow{wmmse_local} - \thisrow{wmmse_local_ci}]{fig2-3.txt};
\addplot [name path=upper, fill=none, draw=none] table [
    x=iteration, y expr=\thisrow{wmmse_local} + \thisrow{wmmse_local_ci}]{fig2-3.txt};
\addplot[orange!10] fill between[of=lower and upper];

\addplot [name path=lower, fill=none, draw=none] table [
    x=iteration, y expr=\thisrow{wmmse_sc} - \thisrow{wmmse_sc_ci}]{fig2-3.txt};
\addplot [name path=upper, fill=none, draw=none] table [
    x=iteration, y expr=\thisrow{wmmse_sc} + \thisrow{wmmse_sc_ci}]{fig2-3.txt};
\addplot[green!10] fill between[of=lower and upper];

\end{axis}
\end{tikzpicture}
}

\caption{In this evaluation, the algorithm's ability to adapt to changes in the network is tested by introducing or removing latent interference sources, while keeping the number of links in $\KK$ constant at 50. The scenario in which no latent interference exists in either the training or testing phase is used as a baseline, as seen in Figure~\ref{fig:zero-zero}. The results are based on 50 independent simulations, with a 90\% confidence interval plotted.}
\end{figure}
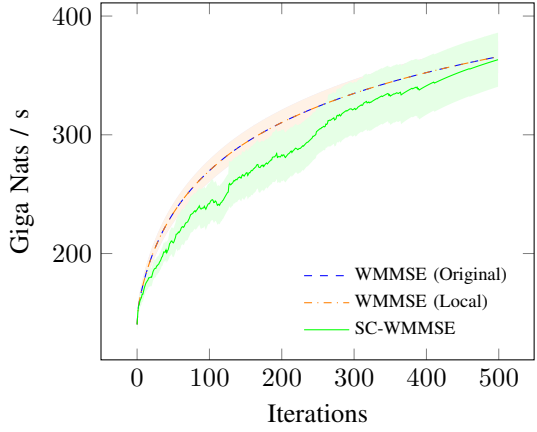
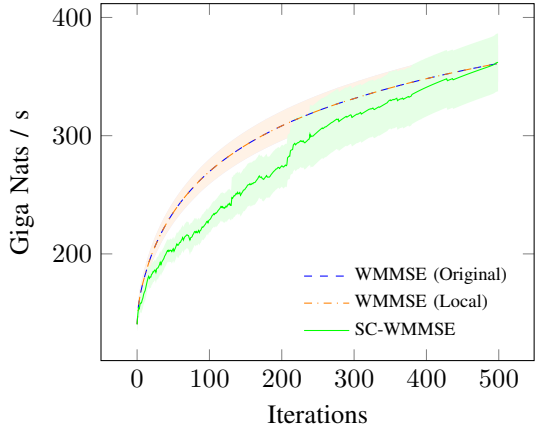
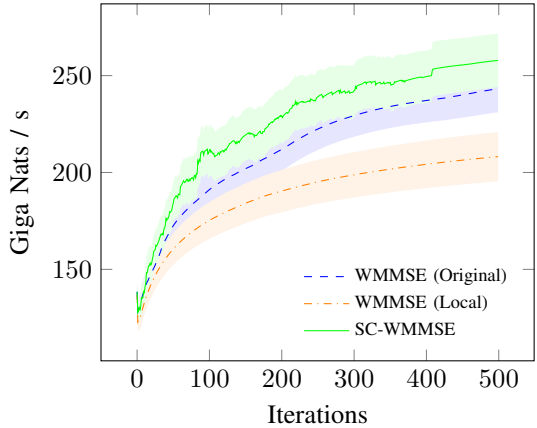

\section{Simulation}
\label{sec:simulation}
The simulation setups are as follows: multiple transmitters are randomly and uniformly distributed within a circle with a radius of $200$ meters. For each transmitter, multiple target receivers are randomly and uniformly distributed within a circle with a radius of $25$ meters. The path-loss model follows the \textit{InH-Shopping Malls-NLOS dual slope} model in \cite{haneda2016indoor}, which takes into account the the probability of line-of-sight shadow fading and blockage. The model applies across 0.5-100 GHz band and 60 GHz is selected for the simulations in this paper. A flat channel is considered and the total bandwidth is $80$ MHz. The maximum power on a \gls{ru} is set to $200$ mW uniformly for all links.

A random matrix $\vec{Z}$ is generated following a uniform distribution $\U(-1,1)$ unless specified otherwise. The power $\vec{q}$ is obtained by the linear transformation $\vec{q}=\vec{Z}\vec{p}$ with randomness and is always capped by the maximum power limit after the transformation\footnote{Note that $\vec{p}$ in the optimization process depends on the iteration step (and hence time), making $\vec{q}$ is actually evolving with time too.}. Each figure in this section is based on $50$ independent simulations to ensure that the results are statistically significant. In every simulation, the network topology is regenerated by following the rules stated above. Additionally, the random matrix $\vec{Z}$, if used, is refreshed in each simulation, for both the training and inference stages.

The code of the simulation is available on \faGithub \cite{papercode}.

\subsection{Objective Performance and Convergence}

This subsection aims to evaluate the effectiveness of the proposed algorithm, SC-WMMSE, in optimizing the objective function in~\eqref{eq:opt} and its performance in terms of convergence. The original \gls{wmmse} algorithm is used as a baseline for comparison, as well as a version of \gls{wmmse} that only uses local network information for optimization, referred to as \gls{wmmse} (Local). In the local version, the term $\eta_{k}$ in Algorithm~\ref{alg:wmmse} is set to $\sigma^2_k$, effectively ignoring all changes in $\KU$. This is used to gauge the impact of $\KU$ on $\KK$ and to determine if such impact is too small to expect SC-WMMSE having an effect.

The results are shown in \figurename~\ref{fig:convergence} with $|\KK|=50$ and $|\KU|$ ranging from $20$ to $50$. SC-WMMSE consistently outperforms the baseline in terms of objective function maximization. Particularly, when the transmissions in the latent network $\KU$ are dense, the algorithm demonstrates significantly better performance in both objective maximization and convergence. On the other hand, when the transmissions in $\KU$ are sparse, \gls{wmmse} (Local) performs similarly to the original \gls{wmmse}. This suggests that the impact of $\KU$ on $\KK$ is low, and the term $\eta_{k}$ can be well approximated by $\sigma_k^2$. As a result, the update in line~\ref{alg:sc-wmmse-estimate} becomes less significant in the optimization process.

\subsection{Robustness}

The robustness of SC-WMMSE in dynamic network changes is evaluated in this section. The algorithm's ability to adapt to emergence or disappearance of interference sources in $\KU$ is in question. \figurename~\ref{fig:zero-zero} is used as a baseline scenario with no latent interference links. SC-WMMSE is not expected to perform better than the others since there is no latent interference. The algorithm's performance is on par with the others in terms of objective value at convergence, but the convergence is slower due to added noise via the convex hull approximation $\hat{I}_k$ ($k\in\KK$). \figurename~\ref{fig:zero-dependent} illustrates SC-WMMSE's robustness in dealing with changes in the network. Even though there is no latent interference present during training, the causal estimator can still make impact in the optimization stage, resulting in SC-WMMSE  outperforming the other two in terms of both objective value and convergence. However, the performance is not as good as seen in \figurename~\ref{fig:dep-dep-k50} due to the more significant data distribution change between training and testing.

\subsection{Scalability}
\begin{figure}[!t]
\centering
\begin{tikzpicture}
\begin{axis}[
    ylabel={Giga Nats / s},
    xlabel={$|\KK|$},
    legend style={font=\scriptsize,draw=none, fill=none},
    legend pos=north east,
    legend cell align={left},
    ytick={1.5, 2.0, 2.5,3.0, 3.5},
]
\addplot [color=blue, mark=square,]
 plot [error bars/.cd, y dir = both, y explicit]
 table[x=links, y=wmmse_mean, y error=wmmse_ci]{fig3.txt};

\addplot [color=orange, mark=triangle,]
 plot [error bars/.cd, y dir = both, y explicit]
 table[x=links, y=wmmse_local_mean, y error=wmmse_local_ci]{fig3.txt};

 \addplot [color=green, mark=o]
 plot [error bars/.cd, y dir = both, y explicit]
 table[x=links, y=wmmse_sc_mean, y error=wmmse_sc_ci]{fig3.txt};

\legend{WMMSE (Original), WMMSE (Local), SC-WMMSE}
 
\end{axis}
\end{tikzpicture}
\caption{Scalability is evaluated for the proposed algorithm over $50$ independent simulations, with $|\KU|=50$. The results are presented with a 99 percent confidence interval, which are visually narrow. The y-axis represents the average throughput per link, which is calculated by dividing the sum-rate achieved on average over $500$ algorithm iterations by the number of links.}
\label{fig:scalability}
\end{figure}
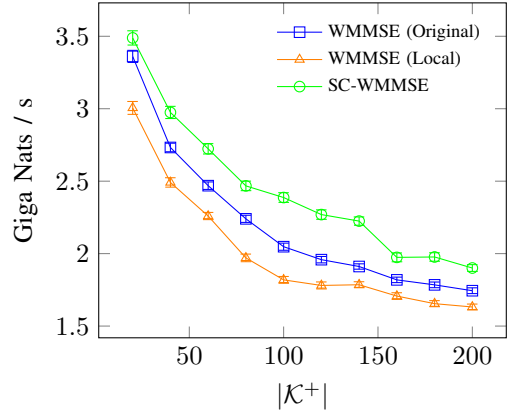

The scalability of the algorithm is evaluated by analyzing its performance on per-link throughput as the density of transmissions in the optimized network, $\KK$, increases. Results show that when the transmissions in $\KK$ are sparse, both \gls{wmmse} and SC-WMMSE perform similarly. However, as the number of links in $\KK$ increases, the advantage of SC-WMMSE becomes more pronounced, until reaching a threshold around 150 links. Beyond this point, the advantage of SC-WMMSE starts to decrease, possibly due to the inherent limitations of the optimization mechanism used by all three algorithms when dealing with large scale problems. It is also worth noting that the gap between \gls{wmmse} and its local version also shrinks with an increase in the number of links in $\KK$. Overall, SC-WMMSE demonstrates better performance than its baseline \gls{wmmse} in all scenarios in \figurename~\ref{fig:scalability}.

\subsection{Necessity of the Convexity Constraint}
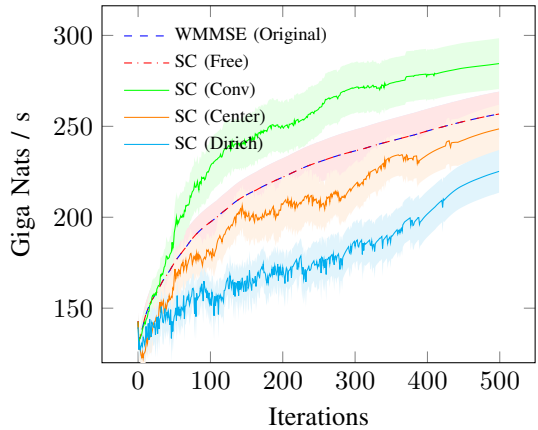
\begin{figure}[!t]
\centering
\begin{tikzpicture}
\begin{axis}[
    ymin=120, 
    xlabel={Iterations},
    ylabel={Giga Nats / s},
    legend style={font=\scriptsize, draw=none, fill=none},
    legend pos=north west,
    legend cell align={left},
    width = 0.4\textwidth,
]
\addplot [blue, dashed] table [x=iteration, y=wmmse]{fig4.txt};
\addplot [red, dash dot] table [x=iteration, y=wmmse_sc_uncons]{fig4.txt};
\addplot [green] table [x=iteration, y=wmmse_sc]{fig4.txt};
\addplot [orange] table [x=iteration, y=wmmse_center]{fig4.txt};
\addplot [cyan] table [x=iteration, y=wmmse_random]{fig4.txt};
\legend{WMMSE (Original), SC (Free), SC (Conv), SC (Center), SC (Dirich)}

\addplot [name path=lower, fill=none, draw=none] table [
    x=iteration, y expr=\thisrow{wmmse} - \thisrow{wmmse_ci}]{fig4.txt};
\addplot [name path=upper, fill=none, draw=none] table [
    x=iteration, y expr=\thisrow{wmmse} + \thisrow{wmmse_ci}]{fig4.txt};
\addplot[blue!10] fill between[of=lower and upper];

\addplot [name path=lower, fill=none, draw=none] table [
    x=iteration, y expr=\thisrow{wmmse_sc_uncons} - \thisrow{wmmse_sc_uncons_ci}]{fig4.txt};
\addplot [name path=upper, fill=none, draw=none] table [
    x=iteration, y expr=\thisrow{wmmse_sc_uncons} + \thisrow{wmmse_sc_uncons_ci}]{fig4.txt};
\addplot[red!10] fill between[of=lower and upper];

\addplot [name path=lower, fill=none, draw=none] table [
    x=iteration, y expr=\thisrow{wmmse_sc} - \thisrow{wmmse_sc_ci}]{fig4.txt};
\addplot [name path=upper, fill=none, draw=none] table [
    x=iteration, y expr=\thisrow{wmmse_sc} + \thisrow{wmmse_sc_ci}]{fig4.txt};
\addplot[green!10] fill between[of=lower and upper];

\addplot [name path=lower, fill=none, draw=none] table [
    x=iteration, y expr=\thisrow{wmmse_center} - \thisrow{wmmse_center_ci}]{fig4.txt};
\addplot [name path=upper, fill=none, draw=none] table [
    x=iteration, y expr=\thisrow{wmmse_center} + \thisrow{wmmse_center_ci}]{fig4.txt};
\addplot[orange!10] fill between[of=lower and upper];

\addplot [name path=lower, fill=none, draw=none] table [
    x=iteration, y expr=\thisrow{wmmse_random} - \thisrow{wmmse_random_ci}]{fig4.txt};
\addplot [name path=upper, fill=none, draw=none] table [
    x=iteration, y expr=\thisrow{wmmse_random} + \thisrow{wmmse_random_ci}]{fig4.txt};
\addplot[cyan!10] fill between[of=lower and upper];

\end{axis}
\end{tikzpicture}
\caption{\gls{sc} (Free) is the unconstrained version of~\eqref{eq:sc}, i.e., \gls{sc} (Conv). Additionally, the evaluation includes two other methods: \gls{sc} Center, which uses the center point of the convex hull for estimation of all units; and \gls{sc} (Dirich), which generates a random Dirichlet distribution as the coefficients $\bm{\beta}$ and selects a random point inside the convex hull in each iteration step.}
\label{fig:necessity_of_convexity}
\end{figure}

The effectiveness of the convex combination constraint in~\eqref{eq:sc-constr} for optimization is demonstrated in \figurename~\ref{fig:necessity_of_convexity}. On top of the synthetic control method used in this paper (the method of solving~\eqref{eq:sc}, referred to as \gls{sc} (Conv)), several other methods are compared, including, its unconstrained version (\gls{sc} (Free)). Additionally,  the center point of the convex hull can be used as the estimate (\gls{sc} (Center)). One could also generate a Dirichlet distribution for the coefficients (\gls{sc} (Dirich)). These methods are used in place of the convex combination constraint in Algorithm~\ref{alg:sc-wmmse} to compare with the baseline \gls{wmmse}.

\gls{sc} (Free) performs similarly to the baseline \gls{wmmse}, which suggests that relaxing the constraint leads to an overfitting to observation, neglecting the counterfactual. Both \gls{sc} (Center) and \gls{sc} (Dirich) perform worse than the baseline, indicating that the inference task is not trivial and that \gls{sc} (Conv) is by-far an effective method for obtaining the coefficients for optimization.

\section{Conclusion and Discussion}

This paper has showcased the potential of using causal inference to assist in solving optimization problems of wireless communications. The proposed SC-WMMSE demonstrates how a causal inference framework can address the challenges posed by confounding variables in optimization. As future work, further analysis and theoretical proofs of the algorithm's convergence are suggested. Additionally, the research in causal inference has typically been based on the assumption of no inter-unit interference, which is not applicable in wireless communications. Therefore, it would be beneficial to develop causal factor models that are better suited for this field.

\bibliographystyle{IEEEtran}
\bibliography{references}

\end{document}